# Terahertz detection of magnetic field-driven topological phase transition in HgTe-based transistors

A. Kadykov,[1,3] F. Teppe,[1,a)] C. Consejo,[1] L. Viti,[2] M. Vitiello,[2] D. Coquillat,[1] S. Ruffenach,[1] S. Morozov,[3] S. Kristopenko,[3] …, M. Marcinkiewicz,[1] N. Dyakonova,[1] W. Knap,[1] V. Gavrilenko,[3] N. N. Michailov,[4] S.A. Dvoretskii[4] …..

[1]*Laboratoire Charles Coulomb (LCC), UMR CNRS 5221, GIS-TERALAB, Universite Montpellier II, 34095 Montpellier, France*

[2]*Institute for Physics of Microstructures, Russian Academy of Sciences, Nizhny Novgorod, Russia*

[3]*NEST, Istituto Nanoscienze-CNR and Scuola Normale Superiore, I-56127 Pisa, Italy*

[4]*Institute of Semiconductor Physics, Siberian Branch, Russian Academy of Sciences, pr. Akademika Lavrent'eva 13, Novosibirsk, 630090 Russia*

We report on Terahertz detection by inverted band structure HgTe-based Field Effect Transistor up to room temperature. At low temperature, we show that nonlinearities of the transistor channel allows for the observation of the quantum phase transition due to the avoided crossing of zero-mode Landau levels in HgTe 2D topological insulators. These results pave the way towards Terahertz topological Field Effect Transistors.

## INTRODUCTION

The main nano-electronics challenges are to reduce power consumption and to improve Field Effect Transistor (FET) operating frequency by reducing their size while retaining or even improving high electron mobility. For these purposes, classical semiconductor technologies are coming to their limits and new materials are explored as possible ways to overcome these restrictions. In this perspective, topological insulators[1] are extremely promising. Whereas their bulk is insulator, their edges are indeed composed of a set of dissipation-less states having Dirac dispersion. Therefore, because their charge carriers are massless and protected from backscattering by the intrinsic topology of the semiconductor band structure, these materials may demonstrate high electron mobility and low power consumption. Moreover, it was shown in different materials that using plasma wave effects in 2D electron channels, nanometer sized FETs can operate as efficient resonant or broadband Terahertz (THz) detectors, mixers, phase shifters and frequency multipliers at frequencies far beyond their fundamental cut-off frequency[2]. Therefore, using plasma wave phenomena in 2D topological insulator-based FETs turns out to be an extremely promising way to take up the present-day nano-optoelectronics challenges.

The notion of topological characterization of condensed matter phases has been introduced in 1982 by Thouless et al.[3]. The quantum Hall (QH) effect was the only known example of such a non-trivial topological phase during 25 years. But in

2005, Kane and Mele have demonstrated that[4] it exists a new phase of matter called quantum spin Hall (QSH) effect[5], in which protected edge states against backscattering can exist without external magnetic field. These metallic states of quantized conductance consist of a pair of 1D helical edge channels moving in opposite directions with opposite spins. In 2006, Bernevig and Zhang have predicted that this QSH phase should occur in HgTe-based quantum wells (QWs) [6,7] when the thickness of the QW exceeds the critical value of 6.3 nm. The existence of two-dimensional (2D) topological insulator (TI) phase in HgTe was experimentally confirmed by Koenig et al.[8]. Unlike classical phase transitions which occur at nonzero temperature due to thermal fluctuation, quantum phase transitions can only occur through changing a physical parameter, such as doping, pressure, or magnetic field. QH and QSH effects are great examples of those quantum phase transitions. Indeed, level crossings in the energy bands of crystals have been identified as a key signature for topological quantum phase transitions. In the 2D TI case, the anticrossing between zero-mode LLs corresponds to the critical magnetic field at which the quantum phase transition appears between QSH insulator and normal insulator states.

In parallel, on the basis of theoretical predictions made by Dyakonov and Shur in 1993[9], Terahertz (THz) photoconductivity using plasma wave nonlinearities in FETs have been demonstrated since more than 10 years in a large number of semiconductor systems[10,11,12,13,14,15]. However, THz experimental techniques were rarely used to probe TI quantum phases and most of the results in the literature concern THz Faraday rotation[16], photogalvanic effect measurements in 3D TIs [17,18] and measurements of the photoelectromagnetic effect induced by THz laser radiation pulses in crystalline TIs[19]. Nevertheless, QH type phases can be efficiently probed by THz radiations. Indeed, HgTe/CdTe based TIs have for example theoretically revealed strong nonlinear optical properties in the THz frequency domain[20]. THz photon energy is also comparable with the Landau level splittings under moderate magnetic fields, and with the topological insulators narrow band gap. Magnetic field and consequently THz radiations are thus powerful tools to probe the properties of these new phases of matter.

In this work, we study an inverted band structure HgTe-based FET up to room temperature by THz photoconductivity as a function of quantizing magnetic field. We show that THz detection by nonlinearities in the FET channel allows for the observation of the magnetic field-induced quantum phase transition between non-trivial inverted band structure and trivial QH regime. These experimental results pave the way towards THz topological FETs.

**Results and discussion**

We have studied an 8 nm wide [013]-oriented HgTe QWs embedded in between $Cd_xHg_{1-x}Te$ barriers with x = 0.7. The sample was grown by molecular beam epitaxy (MBE) on semi-insulating GaAs [013] substrates with relaxed CdTe buffers[21]. Its band structure was inverted. The sample was n-type doped and its dark 2D electrons concentration was ... x$10^{11}$ cm-2.

Photoconductivity experiments were carried out at frequencies of 292 GHz and 660 GHz generated respectively, by a multiplied Schottky diode and a Backward Wave Oscillator (BWO). The THz beam was modulated by a mechanical chopper at 215 Hz and guided to the sample by stainless steel pipes. The transistor was placed at the center of a superconducting coil and cooled to 4.2 K. The detection signal appearing between source and drain under illumination was measured with a standard lock-in technique as a function of the magnetic field at fixed gate voltage.

The FET gate was leaky but allows to slightly controlling the carrier density. Figure 1 shows the electron concentration tuning with the gate bias, extracted from Shubnikov-de Haas (SdH) oscillation frequencies. Electron density varies from approximately 1.1 to 1.8 x $10^{12}$ cm$^{-2}$. Insert a) shows a picture of the FET used in this work. Insert b) illustrates an example of SdH curve obtained up to 16 T at 4.2 K.

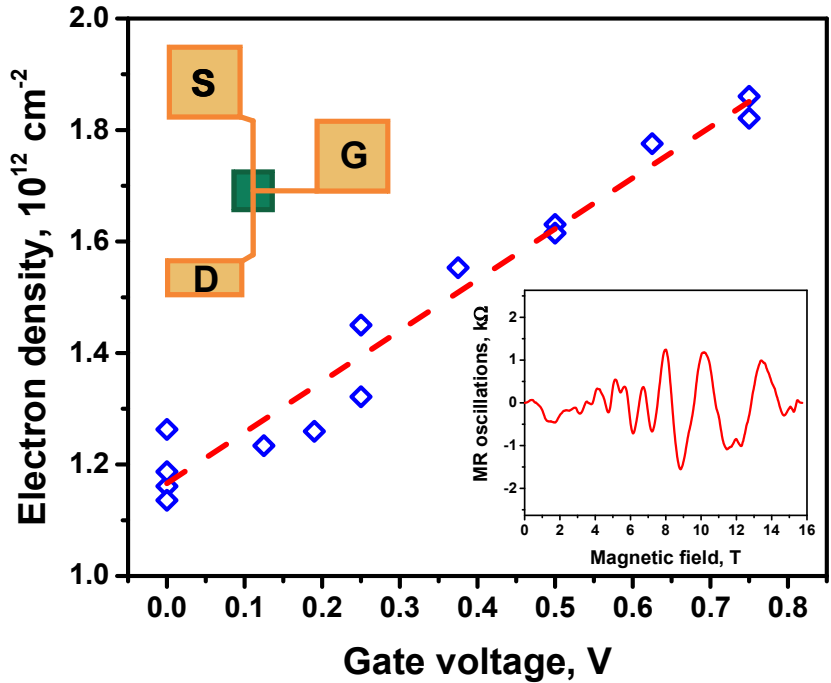

FIG. 1. Measured electron density at 4 K as a function of the swing voltage, extracted from Shubnikov-de Haas oscillation frequencies. Insert a) is a picture of the field effect transistor, when insert b) represents an example of measured Shubnikov-de Haas oscillations up to 16 T.

The FET photoresponse was first measured at 292 GHz as a function of applied perpendicular magnetic field for different gate bias up to 750 mV. In these spectra shown in Figure 2 a), one can clearly see few well pronounced resonances numbered from 1 to …. The first peak corresponds to the cyclotron resonance (CR), as long as it does not move much with the applied gate voltage, but shifts proportionally with the incident frequency from 292 to 660 GHz, giving an effective mass $m^* = 0.03\ m_0$. The particularly interesting resonance 1 positioned at 6.5 T is affected neither by the gate voltage nor by the incident frequency. Unlike the latter, resonances 2 to 7 are well shifting with the carrier density and slightly moving with the incident frequency (see Fig. 2 b)). These unknown resonances cannot be attributed to any magneto-plasmon resonances. Indeed, unlike observed peaks, magneto-plasmon resonant positions must shift towards lower magnetic fields by increasing electron density.

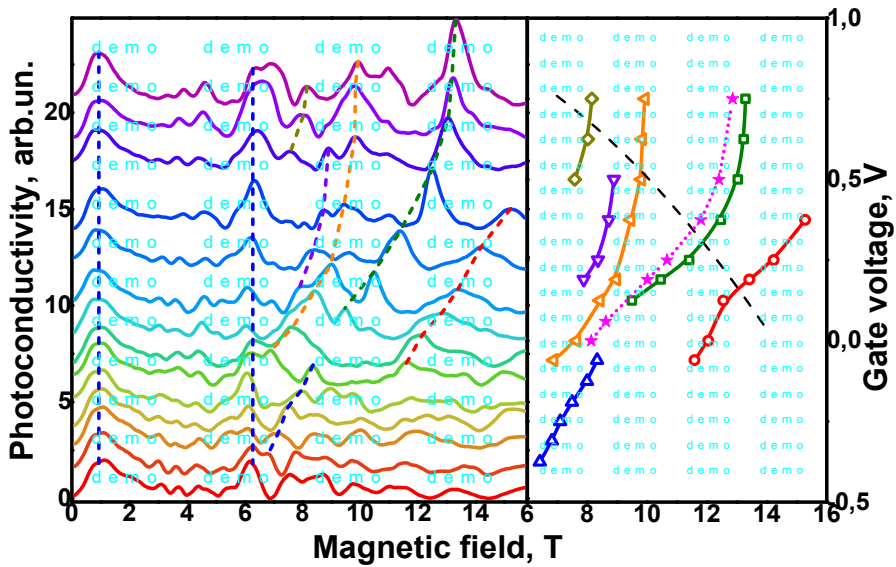

FIG. 2. a) THz photoresponse spectra obtained at 4K at 292 GHz ; b) shows the position of resonances 2 to 7 at 292 and 660 GHz.

When a perpendicular magnetic field is applied, a unique dispersion of the Landau levels is obtained for HgTe inverted band structure QWs. In this regime, the lowest Landau level (LL) of the conduction band contains a pure heavy hole state, when the highest LL of the valence band has a more electronic character and shifts to higher energies in magnetic field. This leads to a crossing of these two peculiar LLs for a finite magnetic field depending on the QW thickness. The observation of such a LL crossing is first a clear indication for the occurrence of an inverted band structure. Therefore, at this critical magnetic field, the Dirac mass parameter M is tuned from negative to positive values, corresponding to a quantum phase transition between inverted, with non-trivial topology, and non-inverted band structure. The crossing of Landau levels for an inverted band structure was demonstrated experimentally by Schultz et al.[22]. However, it was shown in more details in[23] by magneto-transmission Fourier-transform spectroscopy that instead of crossing, these zero-mode LLs are anticrossing[24] because of bulk inversion asymmetry (BIA) of HgCdTe crystal. As measured experimentally and confirmed with the band structure calculations, for an 8 nm wide QW, the LL avoided crossing occurs at approximately 6 T.

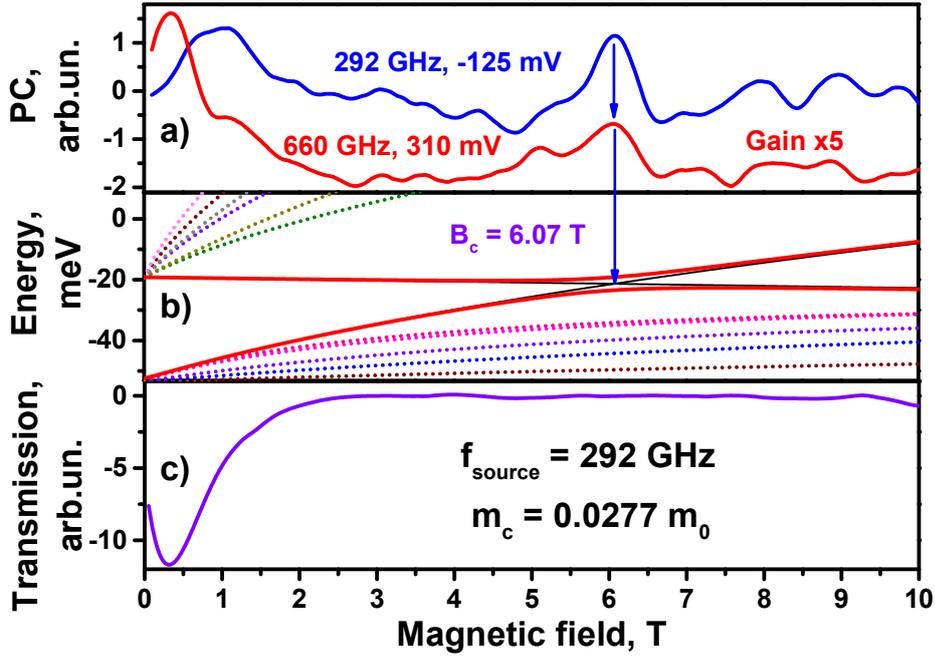

Figure 3 a) shows the photoconductivity spectra as a function of magnetic field at 292 and 660 GHz. At low fields, one can see broad cyclotron resonances shifting with incident frequency. This assumption is in agreement with absorption spectra reported in Figure 3 c) also showing cyclotron resonance lines roughly at the same positions. At approximately 6 T, a well pronounced peak is observed at low temperature in photoconductivity data but interestingly not observed in transmission/absorption spectra. One can notice however that the photon energy used in these experiments is few times larger than the thermal energy in liquid Helium but smaller than the avoided crossing energy, thus can hardly induce optical transition between these two bands. Furthermore, the position of the resonance n°1 is independent on incident frequency and carrier concentration, and corresponds to a critical magnetic field for which particular anticrossing of zero-mode LLs is expected in these samples. Indeed, Kane model calculation of valence and conduction bands splitted into LLs, taking into account BIA, is plotted in Figure 3 b) as a function of magnetic field. One can see the anticrossing of the lower level of conduction band and the upper level of valence band at the critical field $B_c$ = 6 T. Above this field, the non-trivial topological phase is transformed into the conventional quantum Hall phase [8].

It is known that at band structure features, as band edges and anticrossings for example in photonic crystals[25], dielectric nanofilms[26], or microcavities[27], rich dynamics appears. This leads to enhancement of nonlinear optical responses[28] of the material which may be detected by THz photoconductivity technique. Furthermore, THz photoconductivity in field effect transistors (FETs) has been proven since more than 10 years in many materials, to be a powerful tool to probe the 2D electron gaz nonlinearities[29]. V.V. Popov and coworkers have for example studied the theoretical THz photoresponse of a double-quantum-well electron channel of a grid-gated FET. They have shown that the amplitude of the Terahertz electric field increases dramatically in the anticrossing region of optical-like and acoustic-like plasma resonances[30]. On the contrary of transmission spectroscopy, in the photoconductivity experiments, elementary excitations are detected by measuring the photo-induced nonlinear change of the resistance, which monitors exactly the electronic system nonlinearities. This provides chances for exploring unique natures of excitations unable to be investigated by conventional absorption spectroscopy[31]. Therefore, we assume that this photoconductivity signal at approximately 6 T is linked with the quantum phase transition due to the avoided crossing of zero-mode Landau levels in HgTe 2D topological insulators.

**CONCLUSION**

In conclusion, we present the experimental evidence of the room temperature THz detection by inverted band structure HgTe FETs. We have also demonstrated the presence of a well pronounced peak at low temperature and high magnetic field. Unlike other unknown resonances, its position is independent on incident frequency and carrier concentration, and corresponds to a critical magnetic field for which particular anticrossing of zero-mode Landau levels is expected in these samples. Therefore, we assume that this resonance is related to the magnetic field-induced quantum phase transition appearing in HgTe inverted band structure QWs. Other resonances are still not understood and a rigorous qualitative explanation of the observed dependencies requires further theoretical and experimental work.

**ACKNOWLEDGMENTS**

This work was supported by the CNRS through LIA TeraMIR project and by the Russian Academy of Sciences, the non-profit Dynasty foundation, the Russian Foundation for Basic Research (Grants 13-02-00894, 15-02-08274). …

**REFERENCES**

[1] C. L. Kane, J. E. Moore, Physics World 24: 32, (2011).

[2] W. Knap, S. Rumyantsev, M.S. Vitiello, et al. Nanotechnology, V. 24, pp.214002, (2013)

[3] D.J. Thouless, et. al., Phys. Rev. Lett. 49 405, (1982)

[4] C.L. Kane and E.J. Mele, Phys. Rev. Lett. 95, 146802 (2005)

[5] M. Z. Hasan and C. L. Kane, Rev. Mod. Phys. 82, 3045 (2010)

[6] B. A. Bernevig, T. L. Hughes, and S.-C. Zhang, Science 314, 1757 (2006)

[7] A. Roth, C. Brune, H. Buhmann, L. W. Molenkamp, et al., Science 325, 294 (2009)

[8] M. Koenig, S. Wiedmann, C. Brune, A. Roth, H. Buhmann, L. Molenkamp, et al., Science 318, 766 (2007)

[9] M. I. Dyakonov and M. S. Shur, Phys. Rev. Lett. 71, 2465 (1993).

[10] W. Knap, Y. Deng, S. Rumyantsev, and M. S. Shur, Appl. Phys. Lett. **81**, 4637 (2002).

[11] F. Teppe, W. Knap, D. Veksler, M. S. Shur, A. P. Dmitriev, V. Yu. Kachorovskii, and S. Rumyantsev, Appl. Phys. Lett. 87, 052107 (2005).

[12] S. Boubanga-Tombet, F. Teppe, D. Coquillat, S. Nadar, N. Dyakonova, H. Videlier, W. Knap, A. Shchepetov, C. Gardès, Y. Roelens, S. Bollaert, D. Seliuta, R. Vadoklis, and G. Valušis, Appl. Phys. Lett. **92**, 212101 (2008).

[13] R. Tauk, F. Teppe, S; Boubanga, D. Coquillat, W. Knap, Y. M. Meziani, C. Gallon, F. Boeuf, T. Skotnicki, C. Fenouillet-Beranger, D. K. Maude, S. Rumyantsev and M. S. Shur Appl. Phys. Lett. 89, 253511, (2006)

[14] M.S. Vitiello, D. Coquillat, L. Viti, D. Ercolani, F. Teppe, A. Pitanti, F. Beltram, L. Sorba, W. Knap, A.Tredicucci, Nano Letters, V.12, pp. 96-101, (2012)

[15] L. Vicarelli, M. Vitiello, D.Coquillat, A. Lombardo, A. C. Ferrari, W. Knap, M. Polini, V. Pelligrini, A.Tredicucci, Nature Materials, V. 11, pp. 865-71, (2012)

[16] R. Valdes Aguilar, A.V. Stier, W. Liu, L. S. Bilbro, D. K. George, N. Bansal, L. Wu, J. Cerne, A. G. Markelz, S. Oh, and N. P. Armitage, Phys. Rev. Lett. 108, 087403 (2012)

[17] J. Hancock, J. L. M. van Mechelen, A. B. Kuzmenko, et al., Phys. Rev. Lett. **107**, 136803 (2011)

[18] A. M. Shuvaev, G. V. Astakhov, G. Tkachov, et al., Phys. Rev. B 87, 121104(R) (2013)

[19] S. G. Egorova, V. I. Chernichkin, L. I. Ryabova, E. P. Skipetrov, L. V. Yashina, S. N. Danilov, S. D. Ganichev, and D. R. Khokhlov, in Proc. Nanophysics and Nanoelectronics, XIX International Symposium, Nizhniy Novgorod, Russia, vol. 2, pp. 697–698 (2015).

[20] Qinjun Chen et al., Appl. Phys. Lett., 101, 211109 (2012)

[21] S. Dvoretsky, N. Mikhailov, Y. U. Sidorov, V. Shvets, S. Danilov, B. Wittman, and S. Ganichev, J. Electron. Mater. 39, 918 (2010).

[22] M. Schultz, U. Merkt, A. Sonntag, U. Rössler, R. Winkler, T. Colin, P. Helgesen, T. Skauli, and S. Lovold, Phys. Rev. B 57, 14772 (1998).

[23] M. Zholudev, F. Teppe, M. Orlita, et al., Phys. Rev. B 86, 205420 (2012)

[24] M. S. Zholudev, F. Teppe, S. V. Morozov, M. Orlita, C. Consejo, S. Ruffenach, W. Knap, V. I. Gavrilenko, S. A. Dvoretskii, N. N. Mikhailov, JETP Letters Volume 100, Issue 12, pp 790-794 (2015)

[25] J. Torres et al., Phys. Rev. B 69, 085105 (2004)

[26] Vadym Apalkov and Mark I. Stockman, Phys. Rev. B 86, 165118 (2012)

[27] O. A. Aktsipetrov et al, Laser Physics, Vol. 14, No. 5, 2004, pp. 685–691

[28] Vadym Apalkov and Mark I. Stockman, Phys. Rev. B 86, 165118 (2012)

[29] M. I. Dyakonov, M. S. Shur, Phys. Rev. Lett. 71, 2465 (1993)

[30] V. V. Popov, G. M. Tsymbalov and N. J. M. Horing, J. Appl. Phys. 99, 124303 (2006)

[31] S. Holland, Ch. Heyn, D. Heitmann, E. Batke, et al., Phys. Rev. Lett. 93, 186804 (2004)